# Complex signal amplitude analysis for complete fusion nuclear reaction products

*Yu.Tsyganov*

*FLNR, JINR, 141980 Dubna, Russia*

## Abstract

A complex analysis has been performed on the energy amplitude signals corresponding to events of Z=117 element measured in the $^{249}$Bk+$^{48}$Ca complete fusion nuclear reaction. These signals were detected with PIPS position sensitive detector. The significant values of pulse height defect both for recoils (ER) and fission fragments[1] (FF) were measured. Comparison with the computer simulations and empirical formulae has been performed both for ER and FF signals.

## 1. Introduction

Recently, more than 35 new nuclides with atomic numbers Z between 104 and 118 have been synthesized at the Dubna Gas-Filled Recoil Separator (DGFRS) [1- 8]. It should be noted that some of these experimental results have been clearly confirmed in independent experiments [7-9, 10] involving the study of the chemical properties synthesized atoms. In order to succeed in detecting the synthesis of super heavy nuclides, one has to pay attention to the following:

● an electromagnetic recoil separator design has to provide not only an acceptable value of the nuclide transportation efficiency but also a significant suppression of the background products;

● the heavy-ion beam intensity has to be high enough to overcome the limited cross-section for fusion followed by the evaporation of neutrons;

● a detection system has to provide a sufficient number of parameters in order to identify a nuclide. In addition, the design of the detection assembly has to provide for the suppression of the background products [11-14].

● rotating target design has to provide stable, non-destructive application at extremely intense heavy ion beam

---

[1] For a last chain nuclide decay of a multi chain event

## 2. Reaction of $^{249}$Bk+$^{48}$Ca→117 + 3,4n

The discovery of a new chemical element with atomic number Z=117 was reported in [15]. The isotopes of $^{293}$117 and $^{294}$117 were produced in fusion reactions between $^{48}$Ca and $^{249}$Bk. The $^{249}$Bk was produced at Oak Ridge National Laboratory (ORNL) through intense neutron irradiation of Cm and Am targets for approximately 250 days in the High Flux Isotope Reactor. The Bk chemical fraction, separated and purified at the Radiochemical Engineering Development Center of ORNL, contained 22.2 mg of $^{249}$Bk, only 12.7 ng of $^{252}$Cf, and no other detectable impurities. Six arc-shaped targets, each with an area of 6.0 cm$^2$, were made at the research Institute of Atomic Reactors (Dimitrovgrad, RF) by depositing BkO$_2$ onto 0.74-mg/cm$^2$ Ti foils to a thickness of 0.31 mg/cm2 of $^{249}$Bk. The targets were mounted on the perimeter of disk that was rotated at 1700 rpm perpendicular to the beam direction. The experiments were performed employing the DGFRS [1] and the heavy-ion cyclotron U-400 at JINR. Evaporation residues passing through the separator were registered by a time-of-flight (TOF) system with detection efficiency 99.9%, and were implanted in a 4 x 12 cm$^2$ PIPS detector array with 12 vertical position sensitive strips surrounded by eight 4 x4 cm$^2$ side detectors. In order to reduce the background rate in the detector, the beam was switched off for several minutes after a recoil signal was detected with parameters of ER energy expected for 117 ERs, followed by an α-like signal with an energy between 10.7 and 11.4 MeV, in the same strip, within 2.2 mm position window [16]. During the irradiation of Bk target six chains attributed to 3n and 4n de-excitation channels were detected. The values of implanted into PIPS detector ER signals were measured as 8.762, 11.89, 13.87, 13.51, 9.96 and 9.36 MeV.

## 3. Amplitude analysis of ER and SF signals detected in the experiment with the PIPS detector

The multi-parameter events corresponding to production and decays of the super heavy elements (SHE) usually consist of the time-tagged recoil signal amplitudes and the α-decay signal amplitudes. The amplitudes of the signals associated with one or two fission fragments might be present as well. The pulse amplitudes of ERs and FF are observed with a significant pulse height defect (PHD); nevertheless, they are also of great interest since their presence at the beginning and end of each decay chain makes the identification process complete. F.P.Hessberger was the first who recognized the importance of such analysis and demonstrated its validity using Monte-Carlo simulation of FF decays of $^{256}$Rf nuclei implanted into a silicon radiation detector [17]. A simulation method for modeling of ER spectra obtained from DGFRS is reported in Refs. [18-22]. ER registered energy spectrum was calculated by a Monte Carlo simulation taking into account neutron evaporation, energy losses in the different media[2], energy stragglings, equilibrium charge states distribution width in hydrogen, pulse height defect in PIPS detector, fluctuations of PHD. The successful application of these techniques to the data generated in an experiment which was carried out to investigate nuclides with atomic number Z=112 has been reported in [11]. In [19] a simple empirical equation was obtained as

$$E_{REG} = -2.05 + 0.73 \cdot E_{in} + 0.0015 \cdot E_{in}^2 - (\frac{E_{in}}{40})^3$$. Here, $E_{in}$ – is an incoming ER energy value

in MeVs and $E_{REG}$ - the registered detector value.

In the Fig.1 simulation reported in [19] for ER Z=118[3] is shown. An agreement between simulation and measured events is evident. Moreover, if to use formulae (1) one can obtain, taking into account 18.14 MeV calculated incoming energy, the calculated registered value as 11.59 MeV, whereas the mean measured value is equal to 11.22±0.89 MeV.

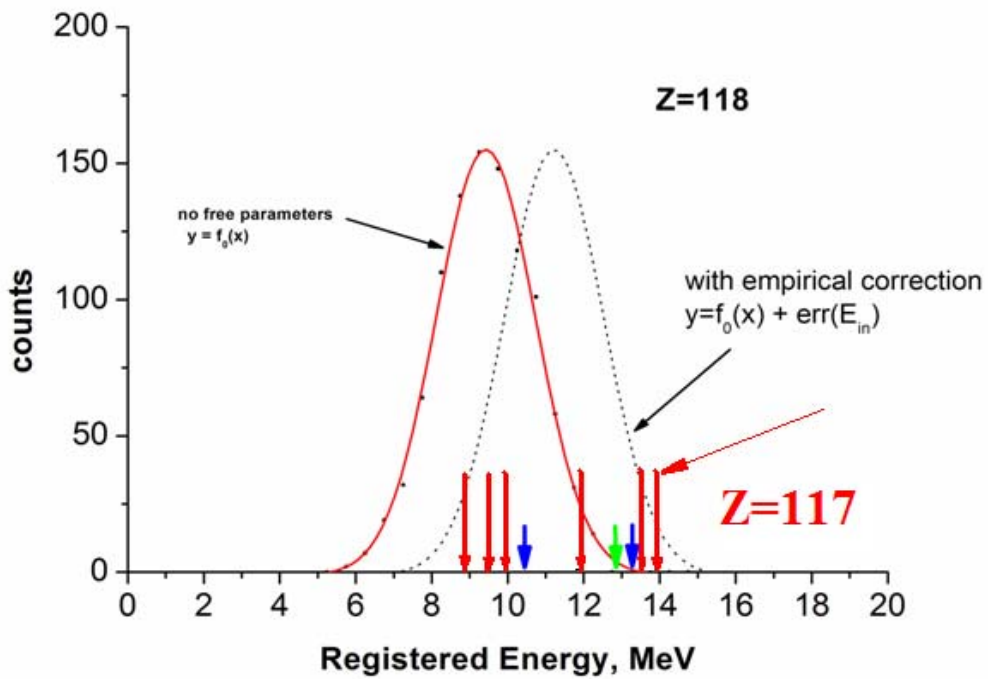

**Fig.1** Computer simulation of Z=118 ER spectrum for $^{249}$Cf+$^{48}$Ca reaction [19]. Amplitudes for Z=117 are shown by long arrows, whereas Z=118 three ER events are shown by short arrows.

Following the philosophy of Ref. [16] a rough mass number estimate could be provided.

---

[2] Target material, hydrogen in the DGFRS volume, Mylar window, pentane in the TOF module
[3] Kinematics close to Z=117 experiment conditions

Namely, taking into account effective shift value between the measured mean value and the registered model spectrum for $^{252}$No recoil (Fig.2) one can calculate as:

$<A> \approx A_0 + h \cdot \delta E$, where $A_0=252$ and $h \approx 13.75$ a.m.u./MeV [16].

Therefore, $<A> \approx 252 + 13.75 \cdot 1.978 = 279.2$ a.m.u.

If one take into account standard deviation[4] of the mean value to be of 0.89 MeV the value of 95% confidence interval should be considered as (255, 303).

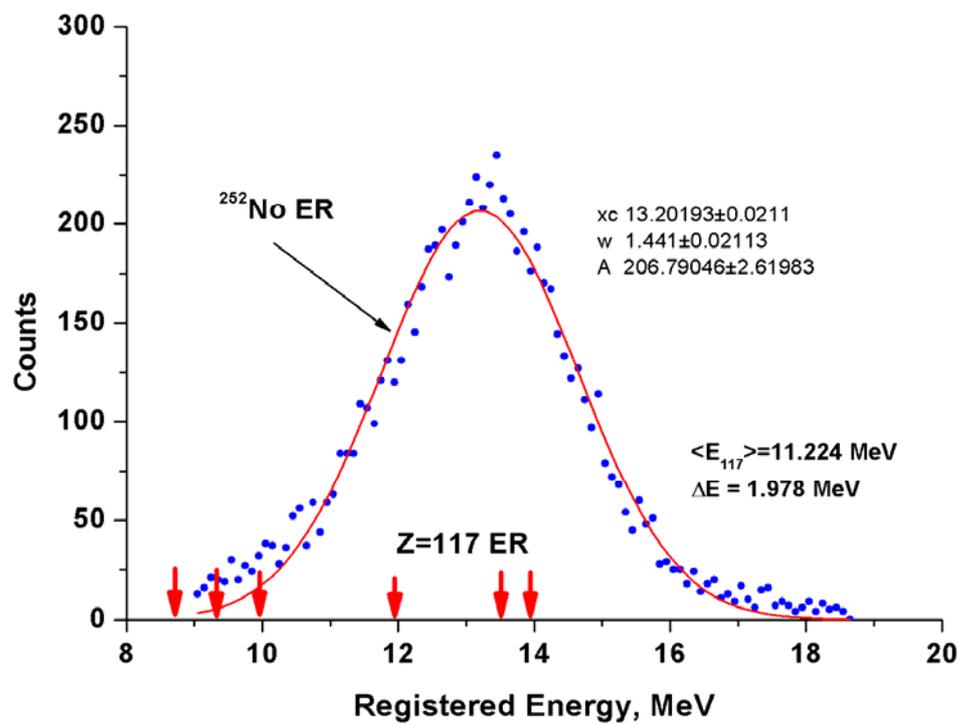

**Fig.2** $^{252}$No ER registered energy model spectrum. Arrows - registered energy Z=117 ER amplitude signals.

As concerns to the registered pair FF signal values one can consider k – parameter systematic [16,23] $k = f(r_{impl})$, where $k = \dfrac{E_{esc}}{E_{foc} + E_{esc}}$. Here energies $E_{esc}$ and $E_{foc}$ are corresponded to the side and focal plane detectors respectively. Result of such representation is shown in the Fig.3. Four events are shown by filled circles at the point of implantation depth of about 3.3 μm.

---

[4] Model systematic error is not taken into account

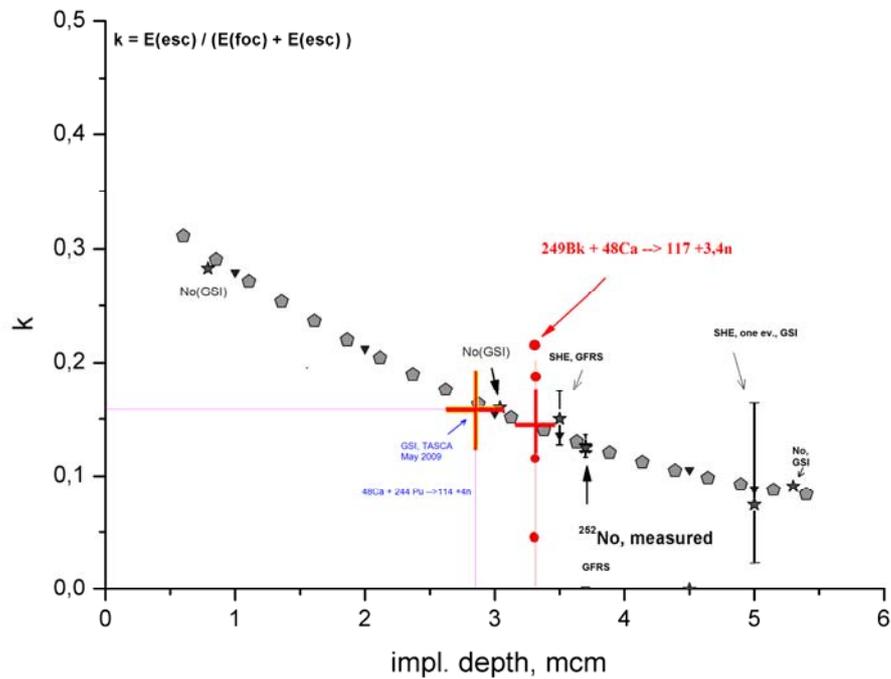

**Fig.3** The dependence of k-parameter against the ER implantation depth.
Pentagons denote calculated values. [10, 19, 23]

## 4. Summary

The complex analysis of the measured energy signal amplitudes in heavy-ion induced complete fusion reactions has been performed for the data from

$^{249}$Bk+$^{48}$Ca $\rightarrow$ $^{294,293}$117 + 3,4n experiment. The experimental data are compared with the results of simulations and empirical equations in this approach. Agreement of these experimental data and numerical ones provide a good independent verification of the experimental results and conclusions of the Ref. [15].

Author plans to continue his effort to develop the approaches aiming at the critical analysis of the experimental data of SHE experiments measured with silicon radiation detector in the nearest future. This paper is supported in part by RFBR grant №09-02-12060.